\documentclass[10.75pt,revtex]{article}
\usepackage[english]{babel}
\usepackage{graphicx}
\usepackage{amsmath}
\usepackage{amsfonts}
\usepackage{amssymb}

\begin{document}

\title{Lorentz Violation on The Primordial Baryogenesis.}
\author{Jorge Alfaro and Pablo Gonz\'alez.\\ \textit{Facultad de F\'isica, Pontificia Universidad Cat\'olica de Chile.}\\
\textit{Casilla 306, Santiago 22, Chile.}\\ jalfaro@uc.cl,
pegonza2@uc.cl}
\maketitle

\section*{ABSTRACT}

Recently many studies have considered the possibility of a Lorentz
  Invariance Violation (LIV), and explored its consequences in a wide range
  of experiments. If this is true, a LIV could explains some mysteries in
  Cosmology. In this paper specifically, we will analyze the effects on The
  Primordial Baryogenesis because it is one of the more important and
  mysterious phenomena of the Big-Bang, that happened at very high energies,
  so we have a real chance to obtain an important effect. We will see that
  this effect could exist, depending directly on the temperature, that is
  very high at this time in the history of the Universe. So, it is possible
  to use this result as a test for a LIV and explore the possibility that the
  boson that started the baryogenesis explains, in part, the dark matter. We
  will obtain estimates about the beginning time of the baryogenesis and the
  boson mass too, that come directly from the LIV.

\section*{INTRODUCTION}

In $1928$ Paul Dirac, in his famous equation, predicted the
existence of anti-particles. Now this has been experimentally
checked: to each particle one anti-particle is associated, where
both have the same mass, half life and opposite charge.  It is
natural to think that they must exist in the same number, but the
reality clearly shows the opposite; it shows an asymmetry
baryon-antibaryon. The production of both and this asymmetry is
named \textit{Primordial Baryogenesis}.\\

What Baryogenesis's theories have in common are the $B$, $C$ and
$CP$ violation, maintaining $CPT$ symmetry, and Departure from
Thermal Equilibrium\footnote{These conditions were proposed for
first time by Sakharov \cite{sajarov}.}. One of the favorite
theories to explain the Baryogenesis consists of introducing a new
particle named \textit{X Boson} that is of the same form as  $Z$
and $W$ bosons, and  should be related to a new force that is
necessary for the \textit{Grand Unification Theory} \cite{Early
uni}.\\

We will study this phenomenon using statistical mechanics and
cosmology, but we will introduce a Lorentz Invariance Violation
(\textit{LIV}) too \cite{liberati}-\cite{scully_stecker}, that is
translated in a small modification of the usual energy dispersion
relation.\\

In the \textbf{Section 1} we will show the usual procedure to represent the Baryogenesis in the most
simple model, explaining in detail the conditions previously
mentioned. In the \textbf{Section 2} we will show the effects when
the LIV is included through Threshold Energy conditions and we will
incorporate them to the development of the previous section.
Finally we will expose the results and some predictions obtained
by the LIV. An extra effect by the LIV is a possible breaking of
\textit{CPT} symmetry, but we will forget it because these effects
turned out to be very little.\\

The LIV that, apparently,
is the most important at high energies, and that we will use here,
has the form \cite{Alfaro2} \cite{Alfaro3} \cite{Alf_Palm1}
\cite{Alf_Palm2} \cite{scully_stecker}:

\begin{eqnarray}
E^2 = v_{max}^2p^2 + m^2c^4
\end{eqnarray}

where $v_{max} = c(1-\alpha)$ is the maximum particle velocity.
$\alpha$ is very little such that it will be marked to high $p$.
Some estimates of this factor $\alpha$, given by theoretical
analysis of the physic of each particle and experimental limit of
some parameters, say that is approximately $10^{-22}$ or
$10^{-23}$. This value will always be little but it can be
different for distinct particles \cite{Alfaro3}
\cite{scully_stecker}.\\

% -----------------------------------------------------------
%                       seccion 1
% -----------------------------------------------------------

\section{BARYOGENESIS.}

For Baryogenesis to be possible, we need a Boson $X$ that
produces baryons and anti-baryons when decaying. Three conditions
have to be also carried out that will allow the present baryonic
asymmetry \cite{sajarov} \cite{Early uni}:

\subparagraph*{a)} Baryon number violation. This mean that, while
the Universe expansion is happening, the factor:

\begin{eqnarray}
B=\frac{n_b - n_{\bar{b}}}{s}
\end{eqnarray}

where $n_b$, $n_{\bar{b}}$ and $s$ are the baryon and anti-baryon
numbers and the entropy per comobile volume, must change. If $B$
violation had not existed, the asymmetry would be only given due
to the initial condition because $B = cte$. So, if we consider a
$X=\bar{X}$ model, there must been a reaction of the kind:

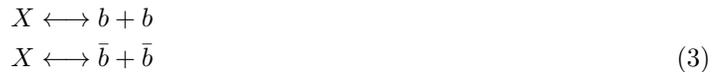
\begin{eqnarray}
X \longleftrightarrow b + b\nonumber \\
X \longleftrightarrow \bar{b} + \bar{b}
\end{eqnarray}

\subparagraph*{b)} C and CP Violation, with CPT symmetry. Of this
form baryons and anti-baryons have the same physics (same mass,
Lorentz violation, etc) and that the boson decay rate can be
different (in a small factor) to the inverse boson decay rate. If
these violations did not exist, baryons and anti-baryons will be
created in the same number, without preference for the matter above
the anti-matter. We can interpret the CP violation, in term of the
amplitude of probability, as:

\begin{eqnarray}
\mid M(X \rightarrow b + b)\mid ^2 - \mid M(b + b \rightarrow
X)\mid ^2 = \epsilon_1 \nonumber \\
\mid M(\bar{b} + \bar{b} \rightarrow X)\mid ^2 - \mid M(X
\rightarrow \bar{b} + \bar{b})\mid ^2 = \epsilon_2
\end{eqnarray}

where $1 \gg \epsilon_i > 0$ allow a little inclination in favor
of the matter. As we also want to keep CPT, we must have:

\begin{eqnarray}
\mid M(X \rightarrow b + b)\mid ^2 = \mid M(\bar{b} + \bar{b}
\rightarrow X)\mid ^2 \nonumber \\
\mid M(X \rightarrow \bar{b} + \bar{b})\mid ^2 = \mid M(b + b
\rightarrow X)\mid ^2
\end{eqnarray}

So, we can deduce the most simple model:

\begin{eqnarray}
\label{ampl_prob}
\mid M(X \rightarrow b + b)\mid ^2 = \mid
M(\bar{b} + \bar{b} \rightarrow X)\mid ^2 =
\frac{1}{2}(1+\epsilon)\mid M_0\mid ^2
\nonumber \\
\mid M(X \rightarrow \bar{b} + \bar{b})\mid ^2 = \mid M(b + b
\rightarrow X)\mid ^2 = \frac{1}{2}(1-\epsilon)\mid M_0\mid ^2
\end{eqnarray}

Where $\mid M_0\mid ^2$ is constant.

\subparagraph*{c)} Departure from Thermal Equilibrium. If
equilibrium exists, we permanently have that $n_b = n_{\bar{b}}$,
and the matter and anti-matter amount would always be the same. A
Departure from Equilibrium process is described by the Boltzmann
equation \cite{Early uni}:

\begin{eqnarray}
\hat{\textbf{L}}[f] = \hat{\textbf{C}}[f]
\end{eqnarray}

Where $\hat{\textbf{L}}$ is the Liouville operator,
$\hat{\textbf{C}}$ is the Collision operator and $f$ is the
distribution function. The Liouville operator corresponds to:

\begin{eqnarray}
\hat{\textbf{L}} = p^{\mu}\frac{\partial}{\partial
x^{\mu}}-\Gamma^{\mu}_{\nu
\rho}p^{\nu}p^{\rho}\frac{\partial}{\partial
p^{\mu}}
\end{eqnarray}

and the Collision operator must consider all possible combinations
of the process. Evaluating:

\begin{eqnarray}
p^{\mu}\frac{\partial f}{\partial x^{\mu}}-\Gamma^{\mu}_{\nu
\rho}p^{\nu}p^{\rho}\frac{\partial f}{\partial p^{\mu}} =
\hat{\textbf{C}}[f]
\end{eqnarray}

Using the $FRW$ metric and considering that $p^{\mu} =
[E,v_{max}\vec{p}]$ and $x^{\mu} = [v_{max}t,\vec{x}]$, is reduced
to:

\begin{eqnarray}
\frac{E}{v_{max}}\frac{\partial f}{\partial t} -
\frac{\dot{R}}{R}v_{max}|\vec{p}|^2\frac{\partial f}{\partial E} =
\hat{\textbf{C}}[f]
\end{eqnarray}

Applying $\frac{g v_{max}}{(2\pi \hbar)^3}\int\frac{d^3p}{E}$, it
is obtained:

\begin{eqnarray}
\frac{g}{(2\pi \hbar)^3}\int \frac{\partial f}{\partial t} d^3p - H(t) \frac{g v_{max}^2}{(2\pi \hbar)^3}\int \frac{p^2}{E}\frac{\partial f}{\partial E}d^3p = \frac{g v_{max}}{(2\pi \hbar)^3}\int\hat{\textbf{C}}[f]\frac{d^3p}{E}
\end{eqnarray}

Where $H(t) = \frac{\dot{R}(t)}{R(t)}$ is the Hubble constant.
Integrating by parts the second term on the left and using the
definition of particles number density $n(t) = \frac{g}{(2\pi
\hbar)^3}\int f d^3p$, we obtain:

\begin{eqnarray}
\frac{\partial n(t)}{\partial t} + 3H(t)n(t) = \frac{g
v_{max}}{(2\pi
\hbar)^3}\int\hat{\textbf{C}}[f]\frac{d^3p}{E}
\end{eqnarray}

Where the term of the right represents the departure from
equilibrium. If it is declared null, it is obtained $n(t) =
n_0 R^{-3}(t)$ that is the case of a evolution in equilibrium.

% -----------------------------------------------------------
%                       seccion 2
% -----------------------------------------------------------

\section{THRESHOLD ENERGY AND COLLISION FACTOR.}

As in this paper we want to see Lorentz Violation effect, we can
have energy levels where the reaction is not produced. To obtain
this energy level, we will use the \textit{Threshold Energy}
\cite{Alf_Palm1} \cite{Alf_Palm2} \cite{coleman}, that consists on
the following:\\

The total energy of a system of many particles is:

\begin{eqnarray}
E = \sum_i E_i\left(|p_i|\right) + \xi_j \left(p_0^j - \sum_i
p^j_i\right)
\end{eqnarray}

where $\xi_j$ are Lagrange multipliers that impose the
conservation of momentum, $i$ level to every particle and $j$
represent the vectorial components. Deriving with respect to
$p^j_i$, to minimize $E$, we obtain:

\begin{eqnarray}
\frac{\partial E_i}{\partial p^j_i} \equiv v_i^j =
\xi_j
\end{eqnarray}

So, $E$ is minimum when the velocity is the same for all
particles. We will name this energy $E_{min} = \sum_i E_i$ where
all particles $i$ have the same velocity. Now, if we have two
particles that collide to produce others, we can write the energy
as:

\begin{eqnarray}
E = E_1\left(\vec{p}-|p_2|\hat{n}\right) + E_2\left(|p_2|\right) +
\chi \left(\hat{n}^2 - 1\right)
\end{eqnarray}

where $\hat{n}$ is normalized vector of $\vec{p}_2$, $\vec{p} =
\vec{p}_1 + \vec{p}_2$ is the total momentum and $\chi$ is other
Lagrange multipliers that impose the normalization of $\hat{n}$.
Maximizing $E$ with respect to $\hat{n}$, we obtain:

\begin{eqnarray}
\hat{n} = \frac{\vec{v}_1|p_2|}{2\chi}
\end{eqnarray}

Since $|\hat{n}| = 1$, we have that $\chi = \pm \frac{v_1|p_2|}{2}$,
so:

\begin{eqnarray}
\hat{n} = \pm \frac{\vec{v}_1}{v_1}
\end{eqnarray}

Evaluating in $E$, we can see that the maximization occurs when
$\hat{n} = - \frac{\vec{v}_1}{v_1}$, so when a frontal collision
happen. This case is represented for $E_{max} = E_a + E_b$
where $a$ and $b$ are particles that collide face to face.\\

Finally, the threshold condition is given by:

\begin{eqnarray}
E_{max} \geq E_{min}
\end{eqnarray}

Since $a$ and $b$ go from one to the other and the particles $i$ go
in the same direction, we have:

\begin{equation}
\label{E_Umbral} E_a + E_b \geq \sum_i E_i~~~~p_a - p_b = \sum_i
p_i
\end{equation}

\subsection{Reactions Allowed Zones}

\subsubsection{$X \longrightarrow b_1 + b_2$}

$b_i$ can be a baryon or anti-baryon. Using (\ref{E_Umbral}):

\begin{eqnarray}
E_X \geq E_{b_1} + E_{b_2} = 2E_{b} \nonumber \\
p_X = p_{b_1} + p_{b_2} = 2p_{b}
\end{eqnarray}

as they are the same kind particle and have the same velocity, we
have that $p_{b_1} = p_{b_2} = p_b$ and $E_{b_1} = E_{b_2} = E_b$.
As a dispersion relation is carried out of the form $E^2 =
v_{max}^2p^2 + m^2c^4$, we have:

\begin{eqnarray}
(v_b^2 - v_X^2)p_X^2 \leq (m_X^2 - 4m_b^2)c^4
\end{eqnarray}

If $v_b > v_X$:

\begin{equation}
p_X \leq \sqrt{\frac{m_X^2 - 4m_b^2}{(v_b^2 - v_X^2)}}c^2~~~~p_b
\leq \sqrt{\frac{m_X^2 - 4m_b^2}{4(v_b^2 - v_X^2)}}c^2
\end{equation}

giving an important superior restriction to momentum. If we use
$m_X \gg m_b$ and $v_b^2 - v_X^2 = (v_b + v_X)(v_b - v_X) \simeq
2c^2\partial \alpha$ con $\partial \alpha = \alpha_X - \alpha_b$,
it is reduced to:

\begin{equation}
\label{cota_X-b}p_X \leq \frac{m_Xc}{\sqrt{2\partial
\alpha}}~~~~p_b \leq \frac{m_Xc}{2\sqrt{2\partial \alpha}}
\end{equation}

If $v_b \leq v_X$, we do not have a bound, since
$p_X^2\textrm{, }p_b^2 \geq 0$.\\

\subsubsection{$b_1 + b_2 \longrightarrow X$}

Doing the same development using (\ref{E_Umbral}), we obtain:

\begin{eqnarray}
E_{b_1} + E_{b_2} \geq E_X \nonumber \\
p_{b_1}-p_{b_2} = p_X
\end{eqnarray}
\begin{eqnarray}
4v_b^2p_{b_1}^2 - 4v_b^2p_Xp_{b_1} + p_X^2(v_b^2-v_X^2)-m_X^2c^4
\geq 0
\end{eqnarray}

Where the approximation $m_X \gg m_b$ was used. The solution, in
$p_{b_1}$, is a parabola with a minimum. So the zeros will give us
the bounds of the reaction. If we define $f(p_{b_1}) =
ap_{b_1}^2+bp_{b_1}+c$, we will see that $a = 4v_b^2$, $b =
-4v_b^2p_X$ and $c = p_X^2(v_b^2-v_X^2)-m_X^2c^4$ where the zeros
are given by:

\begin{equation}
p_{b_1,0} = \frac{-b \pm \Delta}{2a}~~~~\Delta ^2 = b^2 - 4ac
\end{equation}

If $\Delta ^2 < 0$, zeros do not exist and $f(p_b) \geq 0$ is
always carried out without bound. On the other hand, if $\Delta
^2 \geq 0$, a zone exists where the reaction is prohibited.
Evaluating, it is seen that $\Delta ^2 = 16v_b^2E_X^2 > 0$, that
is, a bound exists. The zeros are:

\begin{eqnarray}
p_{b_1,0} = \frac{v_bp_X \pm E_X}{2v_b}
\end{eqnarray}

So, the bounds for this reaction are:

\begin{equation}
\label{cota_b-X_1} p_{b_1} \geq \frac{v_bp_X +
\sqrt{v_X^2p_X^2+m_X^2c^4}}{2v_b}~~~\vee~~~p_{b_1} \leq
\frac{v_bp_X - \sqrt{v_X^2p_X^2+m_X^2c^4}}{2v_b}
\end{equation}

As $p_{b_1}$ and $p_{b_2}$ are related to $p_{b_1} - p_{b_2} =
p_X$, that represent the two particles when directly collide face
to face, so $sign(p_{b_1}) = sign(p_{b_2})$. Comparing:\\

For the first bounds:

\begin{equation}
p_{b_1} \geq \frac{v_bp_X +
\sqrt{v_X^2p_X^2+m_X^2c^4}}{2v_b}~~~~p_{b_2} \geq \frac{-v_bp_X +
\sqrt{v_X^2p_X^2+m_X^2c^4}}{2v_b}
\end{equation}

If $p_X \geq 0$, clearly always $p_{b_1} \geq 0$, but $p_{b_2}
\geq 0$ will only be if $(v_b^2 - v_X^2)p_X^2 \leq m_X^2c^4$. If
$p_X \leq 0$, $p_{b_2} \geq 0$ and $p_{b_1} \geq 0$ only if
$(v_b^2 -
v_X^2)p_X^2 \leq m_X^2c^4$.\\

For the second bound:

\begin{equation}
p_{b_1} \leq \frac{v_bp_X -
\sqrt{v_X^2p_X^2+m_X^2c^4}}{2v_b}~~~~p_{b_2} = p_{b_1} - p_X \leq
\frac{-v_bp_X - \sqrt{v_X^2p_X^2+m_X^2c^4}}{2v_b}
\end{equation}

In this case, if $p_X \geq 0$, always $p_{b_2} \leq 0$, but
$p_{b_1} \leq 0$ will only be if $(v_b^2 - v_X^2)p_X^2 \leq
m_X^2c^4$. The same happens if $p_X < 0$. So, the condition to
$sign(p_{b_1}) = sign(p_{b_2})$ is that:

\begin{equation}
\label{cota_b-X_px}
(v_b^2 - v_X^2)p_X^2 \leq m_X^2c^4
\end{equation}

That corresponds to the decay condition that in $X \longrightarrow
b_1 + b_2$. Analyzing, we can simplify (\ref{cota_b-X_1}),
considering $|p_X| = |p_{b_{max}}| - |p_{b_{min}}|$, in:

\begin{equation}
|p_{b_{max}}| \geq \frac{v_b|p_X| + E_X}{2v_b}~~~~|p_{b_{min}}| \geq \frac{-v_b|p_X| + E_X}{2v_b}
\end{equation}

but $p_{b_1}$ can be in any of both regions. So, the bound can be
simplified to:

\begin{equation}
\label{cota_b-X_Eb} p_{b_1} \geq \frac{-v_b|p_X| + E_X}{2v_b}
\end{equation}

That, together to (\ref{cota_b-X_px}), represent the allowed zone
for the reaction. Now that we have the thresholds, we can do our
calculations.\\

\subsection{Collision Factor and Departure from Equilibrium
Condition}

The Collision Factor is:

\begin{equation}
\frac{g v_{max}}{(2\pi
\hbar)^3}\int\hat{\textbf{C}}[f]\frac{d^3p}{E} = -\int
(2\pi\hbar)^4
\delta^4(p_X-p_{b_1}-p_{b_2})~\Upsilon_{X,b_1,b_2}~d\Pi_1d\Pi_2d\Pi_X
\end{equation}

\begin{eqnarray}
\Upsilon_{X,b_1,b_2} &=& f_X \left(\mid M(X \rightarrow b_1 +
b_2)\mid ^2 + \mid M(X \rightarrow \bar{b}_1 + \bar{b}_2)\mid
^2\right)
\nonumber\\
& & - f_{b_1}f_{b_2} \mid M(b_1 + b_2 \rightarrow X)\mid ^2 -
f_{\bar{b}_1}f_{\bar{b}_2} \mid M(\bar{b}_1 + \bar{b}_2
\rightarrow X)\mid ^2 \nonumber
\end{eqnarray}

Where $f_X$, $f_{b_i}$ and $f_{\bar{b}_i}$ are Boson, baryons and
anti-baryons distribution functions respectively, and
$d\Pi_i=\frac{g_b v_b}{(2\pi \hbar)^3}\frac{d^3p_{b_i}}{2E_{b_i}}$
and $d\Pi_X=\frac{g_X v_X}{(2\pi
\hbar)^3}\frac{d^3p_{X}}{2E_{X}}$. The amplitudes are given by the
model previously mentioned but are cancelled out of zone of
calculated thresholds. This expression follows from assuming that the
distribution of boson, baryons and anti-baryons approach a
Boltzmann distribution of the kind:

\begin{eqnarray}
f_X = e^{-\frac{E_X-\mu_X}{k_BT}} \nonumber \\
f_{b_i} = e^{-\frac{E_{b_i}-\mu}{k_BT}} \nonumber \\
f_{\bar{b}_i} = e^{-\frac{E_{\bar{b}_i}+\mu}{k_BT}}
\end{eqnarray}

Where $\mu_X$, $\mu$ are the boson and baryon chemical potential.
We use this  to simplify the expression of the collision
factor. This is acceptable in the high temperature approximation, that we are using  throughout this work.

We must mention that the boson is not decoupled, since if this is
not like that, the effect on the baryonic asymmetry would be
little for being practically Non-Relativistic ($T \lesssim m_X$).
This is that bosons, baryons and anti-baryons are still in
chemical equilibrium with the thermal bath. So we have that
$\mu_{b_i} = -\mu_{\bar{b}_i} = \mu$.\\

Analyzing the product $f_{b_1}f_{b_2}$, we can see that:

\begin{eqnarray}
f_{b_1}f_{b_2} &=&
e^{-\frac{E_{b_1}+E_{b_2}}{k_BT}}e^{\frac{2\mu}{k_BT}} \nonumber \\
&=& e^{-\frac{E_X}{k_BT}}e^{\frac{2\mu}{k_BT}} \nonumber \\
&=& f_X^{eq}e^{\frac{2\mu}{k_BT}}
\end{eqnarray}

Where $f_X^{eq}$ is the chemical equilibrium boson distribution
($\mu_X = 0$). The same way for the product
$f_{\bar{b}_1}f_{\bar{b}_2}$, we have:

\begin{eqnarray}
f_{\bar{b}_1}f_{\bar{b}_2} =
f_X^{eq}e^{-\frac{2\mu}{k_BT}}
\end{eqnarray}

With these relations and using our probability of amplitudes, we
obtain:

\begin{eqnarray}
\frac{g v_{max}}{(2\pi
\hbar)^3}\int\hat{\textbf{C}}[f]\frac{d^3p}{E} &=& -\int (2\pi
\hbar)^4 \mid M_0\mid ^2 \delta^4(p_X-p_{b_1}-p_{b_2}) \times
\nonumber\\
& & \left[f_X  - f_X^{eq}\right]d\Pi_1d\Pi_2d\Pi_X
\end{eqnarray}

Where we have considered that $\mu \ll k_BT$ and we keep up to first
order in $\mu$ and $\epsilon$. As the term $f_X$ comes from the
boson decay and $f_X^{eq}$ of the inverse decay, they have
different integration ranges. For this, we must separate them in the
integral, so:

\begin{eqnarray}
\label{ec_dif} \frac{\partial n(t)}{\partial t} + 3H(t)n(t) &=&
\frac{g_b^2g_Xv_X}{4v_b(2\pi\hbar)^5} \mid M_0\mid ^2
\left(I_{b_1,b_2,X}^{Neq} - I_{b_1,b_2,X}^{eq}\right) \\
I_{b_1,b_2,X}^a &=& \int \frac{f_X^{a}
\delta^4(p_X-p_{b_1}-p_{b_2})}{2E_{b_1}E_{b_2}E_X} v_b^3
d^3p_{b_1}d^3p_{b_1}d^3p_X \nonumber
\end{eqnarray}

Where $I_{b_1,b_2,X}^{eq}$ and $I_{b_1,b_2,X}^{Neq}$ contain the
distribution with and without equilibrium respectively and $p_i =
\textrm{[$E_i$; $v_{max,i}\vec{p}_i$]}$. We will resolve them in
general form and then we will distinguish them. For this, we use
the relation \cite{decou. neu}:

\begin{eqnarray}
\frac{v_b^3 d^3p_{b_2}}{2E_{b_2}} =
d^4p_{b_2}\delta\left(p_{b_2}^2-m_{b_2}^2c^4\right)\Theta(E_{b_2})
\end{eqnarray}

Then we integrate in $d^4p_{b_2}$ using
$\delta^4(p_X-p_{b_1}-p_{b_2})$, that is simply replacing $p_{b_2}
= p_X-p_{b_1}$, having:

\begin{eqnarray}
I_{b_1,b_2,X}^a = \int
\frac{f^a_X\delta\left((p_X-p_{b_1})^2-m_{b_2}^2c^4\right)\Theta(E_X-E_{b_1})}{E_{b_1}E_X}|\vec{p}_{b_1}|^2\sin(\theta_1)dp_{b_1}d\theta_1d\phi_1d^3p_X
\end{eqnarray}

Now we use the second delta to integrate in $\phi_1$. For this, we
have the identity:

\begin{eqnarray}
\delta(F(\phi_1)) = \sum_{\phi_{1,i}}
\frac{1}{|F'(\phi_{1,i})|}\delta(\phi_1 -
\phi_{1,i})
\end{eqnarray}

With:

\begin{eqnarray}
F(\phi_1) &=& p_X^2+p_{b_1}^2-2p_Xp_{b_1}-m_{b_1}^2c^4 \nonumber \\
&=& m_X^2c^4 - 2E_XE_{b_1} +
2v_bv_X|\vec{p}_{b_1}||\vec{p}_X|\left(\cos(\theta_X)\cos(\theta_1)+\sin(\theta_X)\sin(\theta_1)\cos(\phi_1)\right)
\nonumber \\
&=&
-2v_bv_X|\vec{p}_{b_1}||\vec{p}_X|\sin(\theta_X)\sin(\theta_1)\sin(\phi_1)
\end{eqnarray}

We can see that two values of $\phi_{1,i}$ exist; The first between
$0$ and $\pi$, where $\sin(\phi_{1,i})>0$, and other between $\pi$
and $2\pi$, where $\sin(\phi_{1,i})<0$ and equal in module to the
previous one. So, we evaluate in $|F'(\phi_{1,i})|$ with one of
them and multiply by $2$, obtaining:

\begin{eqnarray}
\delta(F(\phi_1)) &=& \frac{2}{|F'(\phi_{1,1})|}\delta(\phi_1 -
\phi_{1,1})\Theta\left(F'^2(\phi_{1,1})\right) \nonumber \\
I_{b_1,b_2,X}^a &=& 2\int
\frac{f^a_X\Theta(E_X-E_{b_1})\Theta\left(F'^2(\phi_{1,1})\right)}{E_{b_1}E_X|F'(\phi_{1,1})|}|\vec{p}_{b_1}|^2|\vec{p}_X|^2\sin(\theta_1)\sin(\theta_X)dp_{b_1}d\theta_1dp_Xd\theta_Xd\phi_X
\end{eqnarray}

Where $\Theta\left(F'^2(\phi_{1,1})\right)$ appears to assure that
$\cos^2(\phi_1) \leq 1$. Besides, the value of $\phi_{1,1}$
carries out:

\begin{eqnarray}
2\left(E_XE_{b_1}-v_bv_X|\vec{p}_{b_1}||\vec{p}_X|\left(\cos(\theta_X)\cos(\theta_1)+\sin(\theta_X)\sin(\theta_1)\cos(\phi_{1,1})\right)\right)
= m_X^2c^4
\end{eqnarray}

Therefore:

\begin{eqnarray}
|F'(\phi_{1,1})| = \sqrt{a\cos^2(\theta_1) + b\cos(\theta_1) + c}
\end{eqnarray}

With:

\begin{eqnarray}
a &=& -(2v_bv_X|\vec{p}_{b_1}||\vec{p}_X|)^2 \nonumber \\
b &=&
-4v_bv_X|\vec{p}_{b_1}||\vec{p}_X|\cos(\theta_X)(-2E_XE_{b_1}+m_X^2c^4)
\nonumber \\
c &=& (2v_bv_X|\vec{p}_{b_1}||\vec{p}_X|\sin(\theta_X))^2 -
m_X^4c^8 - 4E_X^2E_{b_1}^2 + 4E_XE_{b_1}m_X^2c^4
\end{eqnarray}

So, the integral in $\theta_1$ is reduced to:

\begin{eqnarray}
\int_{-1}^1 \frac{\Theta(ax^2+bx+c)}{\sqrt{ax^2+bx+c}} dx
\end{eqnarray}

With $x=\cos(\theta_1)$. The parabola $ax^2+bx+c$ has a maximum
($a<0$) and the zeros are within $-1$ and $1$. This mean that the
integral interval can be extended to [$-\infty$,$\infty$] without
affecting anything thanks to the Heaviside's $\theta$. Using the relation:

\begin{eqnarray}
\int_{-\infty}^{\infty} \frac{\Theta(ax^2+bx+c)}{\sqrt{ax^2+bx+c}}
dx = \frac{\pi}{\sqrt{-a}}\Theta(b^2-4ac)
\end{eqnarray}

we obtain:

\begin{equation}
\label{integral} I_{b_1,b_2,X}^a = \frac{\pi}{v_b^3v_X}\int
\frac{f^a_X\Theta(E_X-E_{b_1})\Theta(b^2-4ac)}{E_Xp_X}dE_{b_1}d^3p_X
\end{equation}

Where we have used that $E_{b_1} = v_bp_{b_1}$. Analyzing the
second Heaviside's, we can see that its argument is positive if:

\begin{eqnarray}
\frac{E_X - v_Xp_X}{2} \leq E_{b_1} \leq \frac{E_X + v_Xp_X}{2}
\end{eqnarray}

Now we must distinguish the following processes.\\

\subsubsection{$X$ Decay} ($a = Neq$) From (\ref{cota_X-b}), we
have the bound:

\begin{equation}
E_{b_1} \leq \frac{m_Xc v_b}{2\sqrt{2\partial
\alpha}}~~~\vee~~~p_X \leq \frac{m_Xc}{\sqrt{2\partial
\alpha}}~~~~\textrm{If: $v_b > v_X$}
\end{equation}

\begin{equation}
~~~~~~~E_{b_1} < \infty~~~\vee~~~p_X < \infty~~~~~~~~~\textrm{If:
$v_b \leq v_X$}
\end{equation}

But we also must carry out the limits imposed by the Heaviside in
the integral. With a bit of analysis, we can see that the bound
given by the \textit{\textbf{Threshold Energy}} is always greater
than the inferior Heaviside limit. Therefore, the integration
limits are:

\begin{eqnarray}
& \frac{E_X - v_Xp_X}{2} \leq E_{b_1} \leq \frac{E_X + v_Xp_X}{2}&
\\
& \textrm{If:}~(v_b \leq v_X)~\textrm{o}~\left[(v_b > v_X)
\textrm{ and } \left(p_X \leq \frac{m_Xc}{2\sqrt{2\partial
\alpha}}\right)\right]& \nonumber
\end{eqnarray}

\begin{eqnarray}
& \frac{E_X - v_Xp_X}{2} \leq E_{b_1} \leq \frac{m_Xc
v_b}{2\sqrt{2\partial \alpha}}&
\\
& \textrm{If:}~\left[(v_b > v_X) \textrm{ and }
\left(\frac{m_Xc}{2\sqrt{2\partial \alpha}} < p_X <
\frac{m_Xc}{\sqrt{2\partial \alpha}}\right)\right]& \nonumber
\end{eqnarray}

Where we have used that $\partial \alpha \ll 1$. Therefore,
evaluating in (\ref{integral}) when \textbf{$v_b \leq v_X$}, we
obtain:

\begin{equation}
\label{I_Neq_b<x} I_{b_1,b_2,X}^{Neq} (v_b \leq v_X) =
\frac{\pi}{v_b^3}\int \frac{f_X}{E_X}d^3p_X \textrm{ }\textrm{
}\textrm{ }\textrm{ With: $0 \leq p_X \leq \infty$}
\end{equation}

These is almost no difference to the case without Lorentz
Violation. But, if \textbf{$v_b > v_X$} we have:

\begin{eqnarray}
&& I_{b_1,b_2,X}^{Neq} (v_b > v_X) = \frac{\pi}{v_b^3v_X} \times \nonumber \\
&& \left[v_X \int_{A}\frac{f_X}{E_X}d^3p_X +
yv_b\int_{B}\frac{f_X}{E_Xp_X}d^3p_X - \frac{1}{2}
\int_{B}\frac{f_X}{p_X}d^3p_X + \frac{v_X}{2}
\int_{B}\frac{f_X}{E_X}d^3p_X\right]
\end{eqnarray}

with $A \rightarrow (p_X \leq y)$ and $B \rightarrow (y \leq p_X
\leq 2y)$ where $y = \frac{m_Xc}{2\sqrt{2\partial \alpha}}$. We
can see that many extra factors due to the Lorentz violation and his
prohibited energy zones appear. If we call $C \rightarrow (p_X
\geq 2y)$ and consider that in the $B$ and $C$ region have that
$E_X = v_Xp_X$, we obtain:

\begin{eqnarray}
\label{I_Neq_b>x} I_{b_1,b_2,X}^{Neq} (v_b > v_X) &=&
\frac{\pi}{v_b^3} \times
\nonumber \\
&& \left[\int \frac{f_X}{E_X}d^3p_X + \frac{4\pi yv_b}{v_X^3}
\int_{B} f_X dE_X - \frac{4\pi}{v_X^3} \int_{B+C}E_X f_XdE_X
\right]
\end{eqnarray}

Where the integration zone in the first integral extend  to all momenta.\\

\subsubsection{Inverse $X$ Decay} ($a = eq$) In this case, the
bound is given by (\ref{cota_b-X_px}) and (\ref{cota_b-X_Eb}),
so:

\begin{eqnarray}
E_{b_1} \geq \frac{-v_bp_X + E_X}{2}
\end{eqnarray}

with:

\begin{eqnarray}
p_X \leq \frac{m_Xc}{\sqrt{2\partial \alpha}} \textrm{ }\textrm{
}\textrm{ }\textrm{ If: $v_b > v_X$} \nonumber \\
p_X \leq \infty \textrm{ }\textrm{ }\textrm{ }\textrm{ }\textrm{
}\textrm{ }\textrm{ If: $v_b \leq v_X$}
\end{eqnarray}

So, the $E_{b_1}$ limits, considering the Heaviside of the
integral, are:

\begin{equation}
\frac{E_X - v_Xp_X}{2} \leq E_{b_1} \leq \frac{E_X +
v_Xp_X}{2}~~~~\textrm{If: $v_b > v_X$}
\end{equation}

\begin{equation}
\frac{E_X - v_bp_X}{2} \leq E_{b_1} \leq \frac{E_X +
v_Xp_X}{2}~~~~\textrm{If: $v_b \leq v_X$}
\end{equation}

Therefore, if $v_b > v_X$, we have:

\begin{eqnarray}
\label{I_eq_b>x} I_{b_1,b_2,X}^{eq} (v_b > v_X) &=&
\frac{\pi}{v_b^3} \int_{A+B} \frac{f^{eq}_X}{E_X}d^3p_X \nonumber
\\
I_{b_1,b_2,X}^{eq} (v_b > v_X) &=& \frac{\pi}{v_b^3}\left[\int
\frac{f^{eq}_X}{E_X}d^3p_X - \frac{4\pi}{v_X^3} \int_C f^{eq}_X
E_X dE_X\right]
\end{eqnarray}

Where we have used that in the region $C$ the boson is ultra
relativistic. And if $v_b \leq v_X$:

\begin{eqnarray}
\label{I_eq_b<x}
I_{b_1,b_2,X}^{eq} (v_b \leq v_X) = \frac{\pi(v_b +
v_X)}{2v_b^3v_X} \int \frac{f^{eq}_X}{E_X}d^3p_X \nonumber \\
I_{b_1,b_2,X}^{eq} (v_b \leq v_X) = \frac{\pi}{v_b^3} \int
\frac{f^{eq}_X}{E_X}d^3p_X
\end{eqnarray}

Where we use that $\frac{v_X - v_b}{v_X} \ll 1$.\\

\subsection{Differential Equation Solution and Analysis}

The distribution functions inside and outside of equilibrium are
related by $f_X = f^{eq}_X e^{\frac{\mu_X}{k_BT}}$. Now we will
evaluate in (\ref{ec_dif}).

\subparagraph{a)} If $v_b \leq v_X$, we use (\ref{I_Neq_b<x}) and
(\ref{I_eq_b<x}):

\begin{eqnarray}
\frac{\partial n(t)}{\partial t} + 3H(t)n(t) =
\frac{g_b^2g_Xv_X\pi}{4v_b^4(2\pi\hbar)^5} \mid M_0\mid ^2
\left(e^{\frac{\mu_X}{k_BT}} - 1\right)\int
\frac{f^{eq}_X}{E_X}d^3p_X
\end{eqnarray}

\subparagraph{b)} If $v_b > v_X$, we use (\ref{I_Neq_b>x}) and
(\ref{I_eq_b>x}):

$$\frac{\partial n(t)}{\partial t} + 3H(t)n(t) =
\frac{g_b^2g_Xv_X\pi}{4v_b^4(2\pi\hbar)^5} \mid M_0\mid ^2
\times$$
\begin{eqnarray}
\left[\left(e^{\frac{\mu_X}{k_BT}} - 1\right) \int
\frac{f^{eq}_X}{E_X}d^3p_X +
e^{\frac{\mu_X}{k_BT}}\frac{4\pi}{v_X^3} \left(y v_b\int_{B}
f^{eq}_X dE_X - \int_{B+C}E_X f^{eq}_XdE_X\right) +
\frac{4\pi}{v_X^3} \int_C f^{eq}_X E_X dE_X\right]
\end{eqnarray}

As the unique part that depends on the time in $f_X^{eq}$ is the
temperature $T$, if we derive it, we have $\frac{\partial
f_X^{eq}}{\partial t} = - \frac{\partial \beta}{\partial t} E_X
f_X^{eq}$, with $\beta = \frac{1}{k_BT}$. So, deriving the
differential equation and remembering that $n_X^{eq} =
\frac{g_X}{(2\pi\hbar)^3}\int f^{eq}_X d^3p_X$, with $n_X =
e^{\beta\mu_X} n_X^{eq}$, we obtain:

\subparagraph{a)} If $v_b \leq v_X$:

\begin{eqnarray}
\ddot{n}(t) + 3\left[\dot{H}(t)n(t) + H(t)\dot{n}(t)\right] = M(t)
\left[n^{eq}_X(t) - n_X(t)\right] + \mu_X \frac{\partial
\beta}{\partial t} \frac{e^{\beta\mu_X}}{e^{\beta\mu_X} - 1}
\left[\dot{n}(t) + 3H(t)n(t)\right]
\end{eqnarray}

\subparagraph{b)} If $v_b > v_X$:

\begin{eqnarray}
\ddot{n}(t) + 3\left[\dot{H}(t)n(t) + H(t)\dot{n}(t)\right] &=&
M(t) \left[n^{eq}_X(t) - n_X(t) + \frac{4\pi g_X}{v_X^3(2\pi
\hbar)^3} \frac{\partial J}{\partial \beta}\right] \nonumber \\
&+& \mu_X \frac{\partial \beta}{\partial t}
\frac{e^{\beta\mu_X}}{e^{\beta\mu_X} - 1} \left[\dot{n}(t) +
3H(t)n(t)\right]
\end{eqnarray}

With $M(t) = \frac{g_b^2v_X}{16\pi v_b^4\hbar^2} \mid M_0\mid ^2
\frac{\partial \beta}{\partial t}$ and

\begin{eqnarray}
J = e^{\beta \mu_X} \left(y v_b\int_{B} f^{eq}_X dE_X -
\int_{B+C}E_X f^{eq}_XdE_X\right) + \int_C E_X f^{eq}_X
dE_X
\end{eqnarray}

It is the factor that represents the Lorentz violation effect.
Integrating:

\begin{eqnarray}
J = \frac{1}{\beta^2}e^{-\beta y v_b}\left[e^{-\beta y
v_b}\left(2\beta y v_b + 1\right) - e^{\beta \mu_X}\left(\beta y
v_b e^{-\beta y v_b} + 1\right)\right]
\end{eqnarray}

By the high temperature, we know that $\beta y v_b \sim 1$,
moreover if the reactions that  produce the Baryogenesis are
sufficiently fast, we have that $\beta \mu_X \ll 1$. So:

\begin{eqnarray}
J & \sim & \frac{1}{\beta^2} = (k_BT)^2 \nonumber \\
\frac{\partial J}{\partial \beta} & \sim & -\frac{1}{\beta^3} =
-(k_BT)^3
\end{eqnarray}

As $\frac{\partial \beta}{\partial t} = \beta H(t)$, so $M(t) =
\frac{g_b^2v_X}{16\pi v_b^4\hbar^2} \mid M_0\mid ^2 \beta H(t)
\geq 0$. This means that:

\begin{eqnarray}
F(\ddot{n}_X,\dot{n}_X,n_X,\mu_X) \propto - H(t)T^2 \propto
-T^4
\end{eqnarray}

where $F$ is the usual differential equation that represents
Baryogenesis without Lorentz violation (or $v_b \leq v_X$). As the
Baryogenesis temperature is very high (Grand Unification Level),
the Lorentz violation effect, when the Baryogenesis starts, it is
very important; whenever $v_b > v_X$. The effects of this factor
on the solution will be seen in a subsequent work. So far, the
important result is that it is possible to find a trace
of a possible Lorentz violation in the Baryogenesis.\\

Remembering the bound found with the Threshold Energy for the
boson decay, if $v_b > v_X$:

\begin{eqnarray}
p_X \leq \frac{m_Xc}{\sqrt{2\partial \alpha}}
\end{eqnarray}

we can find a limit to the temperature when these reactions start.
For this, we are looking for the temperature to fullfill that:

\begin{eqnarray}
\langle p_X \rangle = \frac{m_Xc}{\sqrt{2\partial
\alpha}}
\end{eqnarray}

For this, we need the relation between average momentum and
temperature. Using a Fermi statistic and $E_X = v_Xp_X$, we
obtain:

\begin{eqnarray}
\langle p_X \rangle = \frac{k_BT\pi^4}{30 c \zeta(3)}
\end{eqnarray}

So, the temperature at the beginning of the Baryogenesis is:

\begin{eqnarray}
k_BT_B &=& \frac{30 \zeta(3) m_Xc^2}{\pi^4 \sqrt{2\partial
\alpha}}
\nonumber \\
\frac{k_BT_B}{m_Xc^2} & \approx & 0.3702 \times 10^{11}
\end{eqnarray}

Where we used $\partial \alpha = 5 \times 10^{-23}$
\cite{scully_stecker}. As the energies are in the Grand
Unification level, it is required that \textrm{$k_BT_B$ or
$m_Xc^2$} $\gtrsim 10^{16}$ [GeV]. If we impose this limit to
$m_X$, we obtain a temperature:

\begin{eqnarray}
k_BT_B \gtrsim 0.3702 \times 10^{27} \textrm{[GeV]}
\end{eqnarray}

That matches a too much early era in the universe (Planck era). On
the other hand, if we impose the limit to $T_B$:

\begin{eqnarray}
m_Xc^2 \gtrsim 2.7012 \times 10^{5} \textrm{[GeV]}
\end{eqnarray}

So, in spite of having an extremely high mass ($m_X \gg m_b$),
these values are far below of the Grand Unification level
(Desert). So, it is possible that the $X$ Boson would be observed
in the LHC because the maximum energies are $\sqrt{s} = 14$ [TeV]
in proton-proton collisions \cite{LHC}.\\

\section*{CONCLUSION.}

As we are at a high energy level, in the Grand Unification scale, we
have a greater possibility to find a LIV effect. In this work, we
saw that this effect really exists, but it becomes more important
if the baryon and boson maximum velocities are related by $v_b >
v_X$. Owing to the fact that we do not know the $X$ boson
properties and difficulty to estimate the $\alpha$ parameter, it
is difficult to know if we are really in this case. But if this is
the case, we would have an important trace of a LIV. Additionally, it is
possible that the modification to the differential equation of the
$X$ boson decay will give some information about the probability
amplitudes. It could be interesting if a LIV could be interpreted
as a parameter of the type of $\epsilon$ in (\ref{ampl_prob}). If
this is possible, the LIV could explain the preference for the
matter over the anti-matter.\\

We estimated a condition for the moment when the Baryogenesis
begun, given by the LIV. This condition tells us that $k_BT_B =
0.262 \times 10^{11} m_Xc^2$. Then  the majority of bosons start
to decay. The condition give us a estimation of the mass or
temperature if we have other method to know one of them. Anyway,
it is a condition that only appears if we impose the LIV.\\

Other point to analyze, that we leave proposed, is the possibility
that the bosons are part of the dark matter. The inferior limit
found to the boson mass is very high yet, but it is very close to
the energy limit obtained by the LHC, so the $X$ boson could be
produced. We must say that this mass limit is obtained only by the
LIV too.\\
\section*{Acknowledgments}

The work of J.A. was partially supported by Fondecyt \# 1060646. PG
acknowledges support from Beca Doctoral Conicyt \# 21080490.

\end{document}